\documentclass[twocolumn,showpacs,preprintnumbers,amsmath,amssymb,prb]{revtex4}

\usepackage{graphicx}
\usepackage{dcolumn}
\usepackage{bm}

\begin{document}


\title{Anisotropic optical conductivities due to spin and orbital
orderings\\ in LaVO$_3$ and YVO$_3$: First-principles studies}

\author{Zhong Fang}

\affiliation{Tokura Spin Superstructure Project (SSS), ERATO, Japan
Science and Technology Corporation (JST), c/o National Institute of
Advanced Industrial Science and Technology (AIST), Tsukuba Central 4,
1-1-1 Higashi, Tsukuba, Ibaraki 305-8562, Japan}

\author{Naoto Nagaosa} 

\affiliation{Correlated Electron Research Center (CERC), AIST Tsukuba
Central 4, 1-1-1 Higashi, Tsukuba, Ibaraki 305-8562, Japan;\\
Department of Applied Physics, University of Tokyo, 7-3-1, Hongo,
Bunkyo-ku, Tokyo 113-8656, Japan}

\author{Kiyoyuki Terakura}

\affiliation{Research Institute for Computational Sciences (RICS),
AIST Tsukuba Central 2, 1-1-1 Umezono, Tsukuba, Ibaraki 305-8568,
Japan}


\date{\today}

\begin{abstract}
The anisotropy of low energy (0$\sim$5eV) optical excitations in
strongly correlated transition-metal oxides is closely related to the
spin and orbital orderings. The recent successes of LDA+$U$ method in
describing the magnetic and electronic structures enable us to
calculate the optical conductivity from first-principles. The LaVO$_3$
and YVO$_3$, both of which have $3d^2$ configuration and have various
spin and orbital ordered phases at low temperature, show distinct
anisotropy in the optical spectra. The effects of spin and orbital
ordering on the anisotropy are studied in detail based on our
first-principles calculations.  The experimental spectra of both
compounds at low temperature phases can be qualitatively explained
with our calculations, while the studies for the intermediate
temperature phase of YVO$_3$ suggest the substantial persistence of
the low temperature phase at elevated temperature.
\end{abstract}

\maketitle

\section{Introduction}

The strong couplings among the spin, lattice and charge degrees of
freedom in transition-metal oxides (TMO) are essentially mediated by
the orbital degree of freedom (ODF), which plays a crucial role in
controlling the phases and various physical
properties~\cite{Orbital,LSMO,JPC}. The direct observation of orbital
structure is difficult, yet several experiments~\cite{X-ray,PCMO,LMO}
have been developed to detect the anisotropy induced by spin and
orbital orderings. Among them, the measurement of anisotropic optical
conductivity~\cite{PCMO,LMO} by using polarized light can provide us
with useful information. The low-energy optical excitations below the
strong O $2p$ to transition-metal $3d$ charge-transfer peak mostly
come from the transition-metal $3d$-to-$3d$ transitions. Those
transitions reflect sensitively the spin and orbital structures of the
system through optical transition matrix elements, and show anisotropy
when spin and orbital are cooperatively
ordered~\cite{Millis,Igor}. Nevertheless, the proper analysis of
experimental anisotropic optical spectra requires the detailed
understanding of electronic and magnetic structures.

Both LaVO$_3$ and YVO$_3$ are insulators with a $Pbnm$ orthorhombic
unit cell with $a\approx b\approx c/\sqrt{2}$ at room
temperature. There are two $3d$ electrons per V$^{3+}$ site.
Synchrotron X-ray and neutron diffraction results~\cite{LVO} suggested
that LaVO$_3$ undergoes a magnetic phase transition at
143K~\cite{Miyasaka2} and a structural phase transition at 140K. The
low temperature phase has the C-type antiferromagnetic (AF) spin
configuration ({\it i.e.}, ferromagnetic (FM) coupling along the
$c$-axis and AF coupling in the $ab$-plane). Due to the Jahn-Teller
distortion, one of the V-O bond is longer than other two in the VO$_6$
octahedron. The longer bond lies in the $ab$-plane and its orientation
alternates for the neighboring V sites in the plane. The stacking of
the longer bonds also alternates along the $c$-axis. Hereafter we call
this kind of distortion the G-type Jahn-Teller distortion.  This low
temperature phase has P2$_1$/$a$ crystal symmetry.

The structural and magnetic phases for YVO$_3$ are quite
complicated~\cite{YVO1,YVO2,YVO3}. With lowering the temperature, this
compound first undergoes a structural phase transition at 200K from a
disordered phase to the G-type Jahn-Teller distorted structure, the
same structure as the low temperature phase of LaVO$_3$. Then the
C-type AF ordering develops at 116 K ($T_{N1}$). With further
lowering the temperature to 77K ($T_{N2}$), another structural and
magnetic phase appears. The low temperature ($<$77K) phase recovers
the $Pbnm$ crystal symmetry with the C-type Jahn-Teller distortion
({\it i.e.}, the longer bonds stack along the $c$-axis in the same
orientation rather than alternately). The magnetic structure of this
low temperature phase is G-type AF structure ({\it i.e.}, AF coupling
both in the $ab$-plane and along the $c$-axis). Strong
temperature-induced magnetization reversal can be observed at $T_{N1}$
and $T_{N2}$~\cite{YVO4}.

Sawada and Terakura studied the electronic and magnetic
structures of LaVO$_3$ and YVO$_3$ by using the full-potential
linearized augmented-plane-wave (FLAPW) method. They first compared
the results~\cite{sawada1} between LDA (local density approximation)
and GGA (generalized gradient approximation), and then applied the
LDA+$U$ method~\cite{LDAU} to LaVO$_3$ ~\cite{sawada2}. They
found that the LDA and GGA are not sufficient, and LDA+$U$ is
necessary in order to predict the low temperature phase correctly.  
Recently, the polarized optical conductivities of LaVO$_3$
and YVO$_3$ were measured by Miyasaka et al.~\cite{Miyasaka} and very
clear anisotropy was observed. To understand the implications of 
those spectra, we developed the plane-wave pseudopotential method based on
LDA+$U$ to calculate the optical conductivity and studied 
the effects of spin and orbital orderings on the anisotropy. The
experimental spectra of both LaVO$_3$ and YVO$_3$ at low temperature
can be qualitatively explained by our calculations, while the studies
for the intermediate temperature phase ($77K<T<116K$) for 
YVO$_3$ suggest the
complication of magnetic and orbital structures.  In Section II of
this paper, we will describe our method, and the results are discussed
in Section III.

\section{Method}

The Vanderbilt type ultra-soft pseudopotential~\cite{PP}
is useful not only for efficient calculations for transition-metal oxides
but also for implementing the LDA+$U$ method to treat effects of
strong correlation.
In the LDA+$U$ method, the strong Coulomb
interaction is explicitly taken into account in the subspace of
localized orbitals through a Harteree-like scheme. The detailed
description of our LDA+$U$ scheme was given in 
Ref.[\onlinecite{JPC}].

The inter-band optical conductivity is calculated from the converged
Kohn-Sham wave functions $|\psi_{n\bf k}\rangle$ and eigen values
$E_n({\bf k})$ by using the following Kubo formula~\cite{Kubo} (in Ry
units):

\begin{eqnarray}
\sigma_{\alpha\beta}(\omega)
&=&-\frac{16}{V}\sum_{\bf k\it n}if_{n\bf k}\sum_{m}
\frac{1}{\omega_{mn}^2-(\omega+i\delta)^2} \nonumber\\
&&\left[\frac{\omega+i\delta}{\omega_{mn}}
Re(\pi_{nm}^\alpha\pi_{mn}^\beta)+iIm(\pi_{nm}^\alpha\pi_{mn}^\beta)
\right]
\end{eqnarray}
where $\alpha$ and $\beta$ (=$x,y,z$) are indices for directions,
$\omega$ is the excitation energy, $V$ is the volume of the unit cell,
$n$ and $m$ are band indices, $f_{n\bf k}$ is the Fermi distribution
function, $\omega_{mn}=E_{m}({\rm k})-E_{n}({\rm k})$ and $\delta$ is
the lifetime broadening ($\delta$=0.01Ry in this work),
$\pi_{nm}^\alpha=\langle \psi_{n\bf k}|(-i\nabla_\alpha)|\psi_{m\rm
k}\rangle$ are the matrix elements of the momentum operator.  The
calculations for the matrix elements $\pi_{nm}^\alpha$ require the all
electron wave-functions $\psi_{n\bf k}$, which can be obtained from
the following core compensation form~\cite{Fujiwara,Matrix}:

\begin{equation}
|\psi_{n\bf k}\rangle=|\phi_{n\bf k}\rangle+\sum_{i}
\{|\psi_i\rangle-|\phi_i\rangle\}\langle\beta_i|\phi_{n\bf k}\rangle
\end{equation}
where $\phi_{n\bf k}$ are the pseudo-wave-functions obtained from the
self-consistent pseudopotential calculations, $i$ is the index for
atomic orbitals, $\psi_i$ and $\phi_i$ are atomic all-electron and
pseudo- wave functions respectively, $\beta_i$ are the localized
functions as defined in Ref.[\onlinecite{PP}]. Therefore, the matrix
elements of momentum operator can be obtained
as~\cite{Fujiwara,Matrix}:

\begin{eqnarray}
\pi_{nm}^\alpha&=&\langle \psi_{n\bf k}|(-i\nabla_\alpha)|\psi_{m\bf
k}\rangle \nonumber\\
&=&\langle \phi_{n\bf k}|(-i\nabla_\alpha)|\phi_{m\bf
k}\rangle
+\sum_{ij}\langle\phi_{n\bf k}|\beta_i\rangle
\{ \langle \psi_{i}|(-i\nabla_\alpha)|\psi_{j}\rangle \nonumber \\
&&-\langle \phi_{i}|(-i\nabla_\alpha)|\phi_{j}\rangle \}
\langle\beta_j|\phi_{m\bf k}\rangle
\end{eqnarray}

Practically, the core contribution (second term of eq.(3)) can be
calculated in the pseudopotential generation process and stored as
input of self-consistent calculations. The present process has been
well checked by using LaMnO$_3$ as an
example~\cite{Millis,Igor}. Throughout the calculations, 30Ry has been
used for the cutoff energy of plane-wave expansion, and we use
$(6\times6\times4)$ mesh for the k-points in the linear tetrahedron
method with the curvature correction. The parameter $U_{eff}$ in
LDA+$U$ scheme is chosen to be about 3.0eV in order to
reproduce the experimental band gaps~\cite{gap,Miyasaka,JPC}.

\section{Results and Discussions}
\subsection{Electronic structures}

\begin{figure}
\includegraphics[scale=0.3]{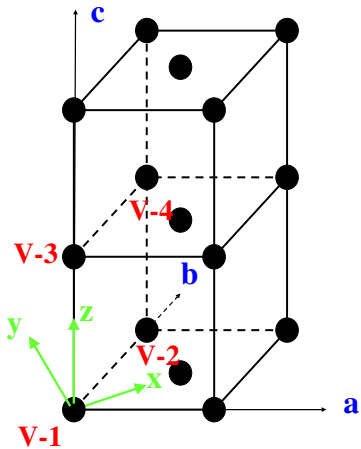}
\caption{The unit cell and coordinates used in the calculations. Four
V sites (black spheres), labeled as V-1, V-2, V-3 and V-4, are
included in the unit cell. }
\end{figure}

The unit cell and coordinates are defined in Fig.1.  Four V atoms, say
V-1, V-2, V-3 and V-4, are included in the unit cell. Following the
common convention, we defined the $x$, $y$ and $z$ directions as the
[110], [\={1}10] and [001] directions of the unit cell
respectively. Three structures, LaVO$_3$ at 10K~\cite{LVO}, YVO$_3$ at
65K~\cite{YVO3}, and YVO$_3$ at 100K~\cite{YVO3} are treated here,
which have the G-type, C-type and G-type Jahn-Teller distortions,
respectively. For each fixed structure, relative stability is studied
among four kinds of collinear magnetic structures: C-type AF, G-type
AF, A-type AF({\it i.e.}, FM layers coupled antiferromagneticaly along
the $c$-axis), and FM spin ordering (SO) states.

Each of the above Jahn-Teller distortions is accompanied by the same
type of orbital ordering (OO) and stabilizes a specific related
magnetic structure.  In the present case, our calculations suggest
that the G-type and C-type OO favor the C-type and G-type AF SO,
respectively.  This result is consistent with experimental
observations and also with the previous
calculations~\cite{sawada1,sawada2} and unrestricted Hartree-Fock
studies~\cite{Mizokawa}.  Now we concentrate on the discussion for the
electronic structures of the lowest energy magnetic state of each
structure. They are A) LaVO$_3$ 10K with G-OO and C-SO; B) YVO$_3$ 65K
with C-OO and G-SO; C) YVO$_3$ 100K with G-OO and C-SO. The obtained
magnetic moments for these three states are all about
1.7$\mu_B$/V. The calculated electronic projected densities of states
(PDOS) are summarized in Fig.2, where different orbitals are shown by
different colors. The O-$2p$ and V-$e_g$ states are not plotted.

\begin{figure*}
\includegraphics[scale=0.7,angle=-90]{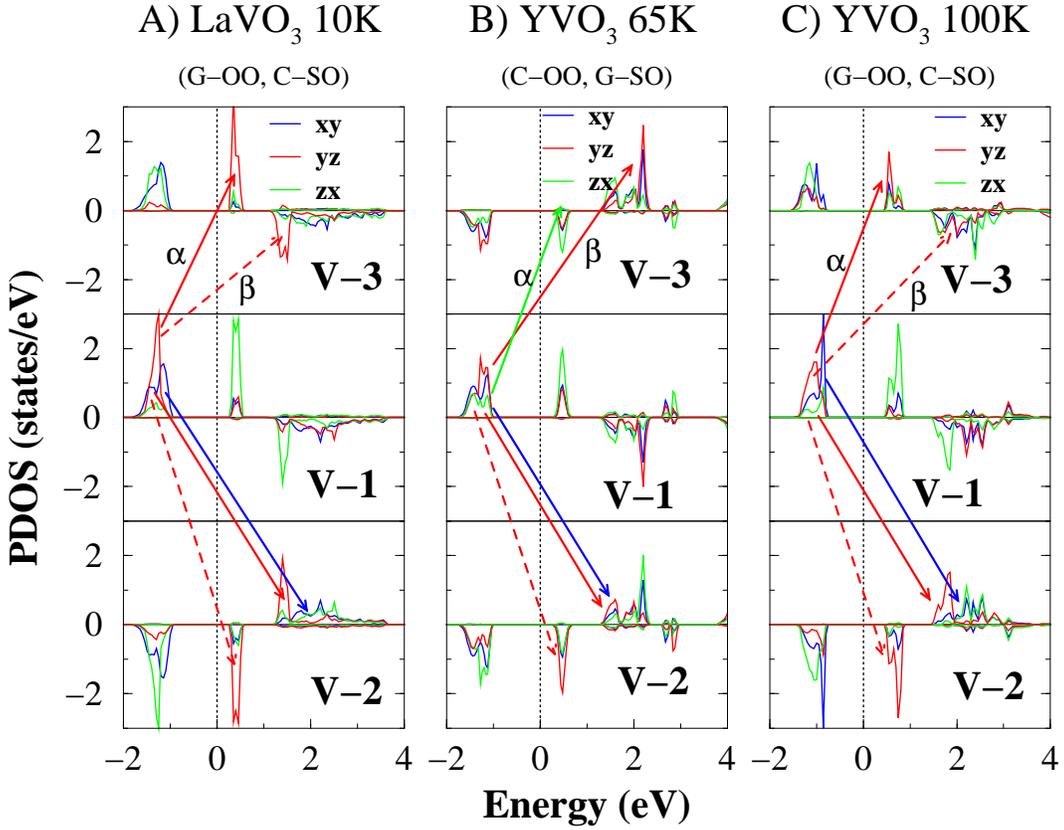}
\caption{The calculated projected densities of states (PDOS) for the
lowest energy states of three systems. They are A) LaVO$_3$ 10K
structure with G-OO and C-SO; B) YVO$_3$ 65K structure with C-OO and
G-SO; C) YVO$_3$ 100K structure with G-OO and C-SO. The lines with
arrow indicate the optical transition paths from V-1 site to other
sites. All the PDOS and paths for different orbitals are indicated by
different colors. The solid arrows show the real transition in this
spin and orbital configuration, while the dashed arrows show the
possible transition in different magnetic structures. The positive and
negative PDOS values mean $\uparrow$-spin and $\downarrow$-spin
respectively. See the text for detailed explanations.}
\end{figure*}

For A), the occupied electronic configuration can be nominally
expressed as V-1: $d_{xy}^{\uparrow}d_{yz}^{\uparrow}$; V-2:
$d_{xy}^{\downarrow}d_{zx}^{\downarrow}$; V-3:
$d_{xy}^{\uparrow}d_{zx}^{\uparrow}$; V-4:
$d_{xy}^{\downarrow}d_{yz}^{\downarrow}$. One of the two electrons on
each V site will occupy the $d_{xy}$ orbital, and another electron will
occupy the $d_{yz}$ or $d_{zx}$ orbital alternately from one V atom
to its neighboring V atoms, resulting in G-OO. The C-SO is manifested as the
up spin state for V-1 and V-3, and the down spin state for V-2 and
V-4. Due to deviation from the cubic structure, certain degree of mixture
of orbitals can be observed: for example, the small occupation of
$d_{zx}^{\uparrow}$ orbital (green line) for V-1 site. Note
here that, in our calculation, for each fixed structure and spin
configuration the orbital polarization is fully relaxed
self-consistently.  The calculated electronic and magnetic structures
of YVO$_3$ in the intermediate temperature (100K) phase (panel C in
Fig.2) can be understood in the same way as the low temperature phase
of LaVO$_3$ in terms of the same spin and orbital orderings. This can
be seen from the similarity of the panel A) to the panel C) in Fig.2.

For the low temperature phase of YVO$_3$ (panel B in Fig.2), the
situation is different. The nominal electronic occupations can be
understood as V-1: $d_{xy}^{\uparrow}d_{yz}^{\uparrow}$; V-2:
$d_{xy}^{\downarrow}d_{zx}^{\downarrow}$; V-3:
$d_{xy}^{\downarrow}d_{yz}^{\downarrow}$; V-4:
$d_{xy}^{\uparrow}d_{zx}^{\uparrow}$. As for the OO and SO in
the $ab$-plane, the present situation is the same as those of A) and C). 
However, the
OO and SO along the $c$-axis are different.  In the present case, 
all the V atoms along a given $c$-axis have the same orbital occupation, 
while those belonging to the nearest neighbor $c$-axes have 
different orbital occupations.
For instance, the $d_{yz}$ orbital is
occupied for both V-1 and V-3 sites. 
This C- OO leads to stability of the G-SO through AF superexchange.  
Comparing LaVO$_3$ with YVO$_3$, we
can find strong mixture of orbitals in YVO$_3$. This is due to the
enhanced structural distortion in YVO$_3$ compared with LaVO$_3$. The
V-O-V angles are about 157 degrees in LaVO$_3$, while they are reduced to
about 144 degrees in YVO$_3$.

Similar analysis was made for all other magnetic phases of 
these three structures.  Among several results obtained, we noticed
the following interesting and possibly important feature: for a fixed
crystal structure, a change in the magnetic structure will not cause any
significant change in the orbital ordering. It is likely that 
the orbital structure
is mostly determined by the specific structure distortion
in the present case.

\subsection{Optical conductivities}

\begin{figure*}
\includegraphics[scale=0.6,angle=-90]{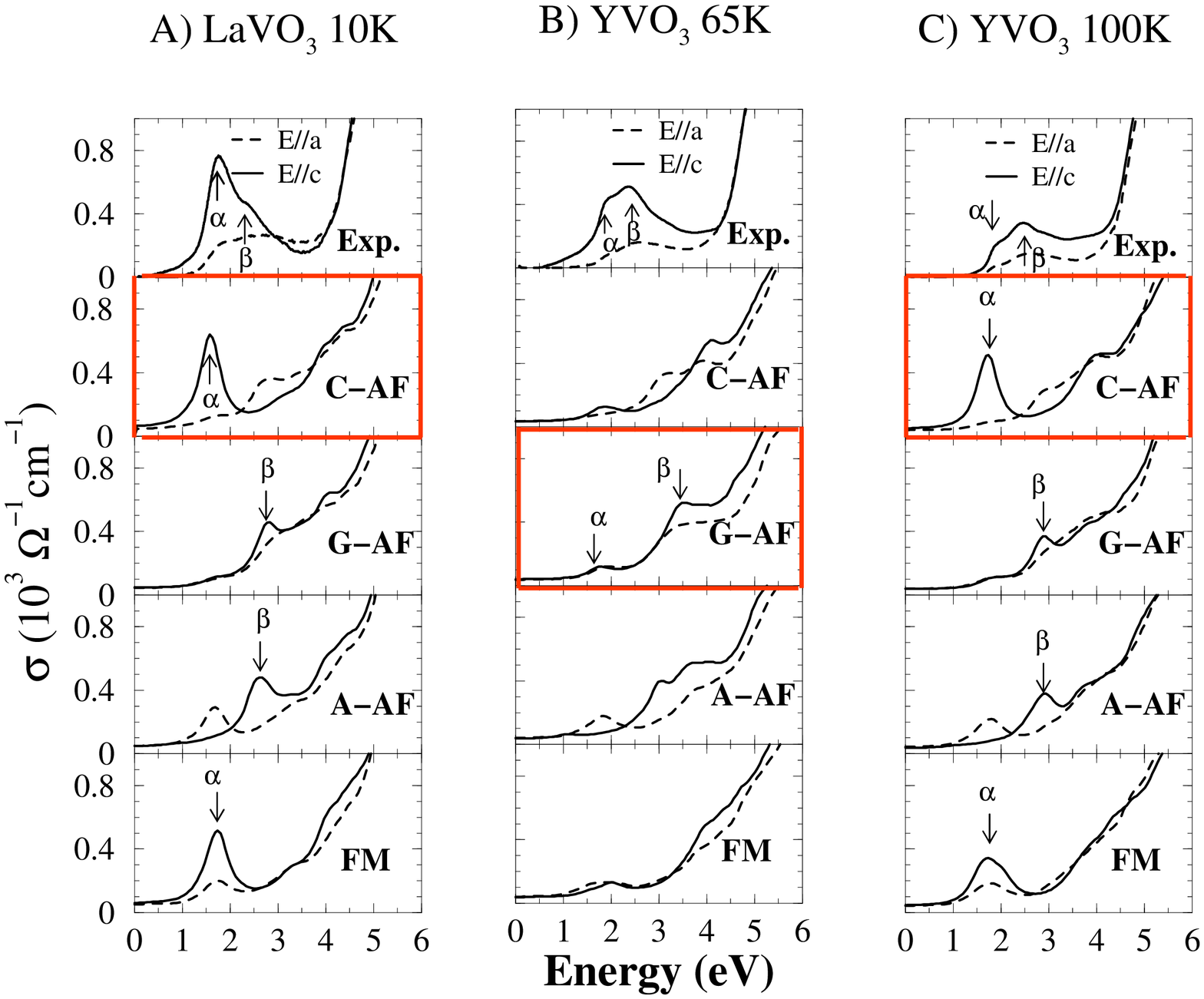}
\caption{The calculated optical conductivities for various
phases. Three structures, A) LaVO$_3$ at 10K, B) YVO$_3$ at 65K, and
C) YVO$_3$ at 100K, are calculated. For each fixed structure, results
for four different magnetic configurations are plotted together with
experimental results. The figure enclosed by the red border line
corresponds to the most stable spin structure. The experimental curves
in A), B) and C) are taken at 10K, 67K, and 110K
respectively~\cite{Miyasaka}. The solid lines are results for E//$c$,
while dashed lines for E//$a$. See the text for the detailed
explanations.}
\end{figure*}

The results for optical conductivity are summarized in
Fig.3. Experimentally, two peak structures ($\alpha$ and $\beta$) are
observed for E//$c$ in both compounds, with peak $\beta$ located at
higher energy than peak $\alpha$. For LaVO$_3$, peak $\alpha$ has a
very sharp structure at low temperature and is significantly
suppressed above the transition temperature. Peak $\beta$ is quite
weak and almost temperature independent. In YVO$_3$, on the other
hand, peak $\beta$ has a larger weight with strong temperature
dependence and peak $\alpha$ is weak with only little temperature
dependence. For both LaVO$_3$ and YVO$_3$, the E//$a$ spectra are
quite broad and almost temperature insensitive.

\subsubsection{LaVO$_3$ low temperature phase:}

For LaVO$_3$ at 10K, the G-OO with C-SO state is the most stable
state.  We first analyze the case of E//$c$, where we consider the
inter-site transition between V-1 and V-3 as the transition dipole is
along the $c$-axis.  The calculated optical conductivity for C-AF in
Fig.3 shows clearly a sharp peak corresponding to the experimental
peak $\alpha$.  Looking back to the calculated PDOS in Fig.2, and
taking the transitions from V-1 site to other sites as examples, we
can assign the sharp peak $\alpha$ for E//$c$ to the transition from
V-1-$d_{yz}^\uparrow$ to V-3-$d_{yz}^\uparrow$ as denoted by the solid
red arrow in Fig.2. This transition is not allowed if the magnetic
structure along $c$-axis is AF, like in the G-AF and A-AF states,
where no peak is obtained around the same position of peak
$\alpha$. Therefore, the peak $\alpha$ should be suppressed above the
magnetic transition temperature as observed in experiment.  On the
other hand, peak $\beta$ seen in experiment does not exist in the
calculated spectra for C-AF.  Since the spectral weight of peak
$\beta$ is very small and temperature insensitive in experiment, we
speculate that this peak may come from certain component of imperfect
spin ordering along the $c$-axis.  For example, let us reverse the
spin of V-3 site in Fig.2A.  The transition from V-1-$d_{yz}^\uparrow$
to V-3-$d_{yz}^\uparrow$ located at higher energy (denoted by the red
dashed arrows in Fig.2) will be allowed and contribute to a sharp peak
near the position of peak $\beta$, as shown in the spectra for the
G-AF and A-AF state for E//$c$.  Before moving to YVO$_3$, we briefly
discuss the case of E//$a$, where the intersite electronic transitions
between V-1 and V-2 are important.  The two transitions from
V-1-$d_{xy,yz}^\uparrow$ to V-2-$d_{xy,yz}^\uparrow$ (denoted as solid
blue and red arrows in Fig.2) are allowed for the $ab$-plane spectra,
which show a broad structure due to the broad band width of unoccupied
minority spin states. These $ab$-plane transitions produce a structure
in E//$a$ curve at the position of peak $\beta$ of E//$c$ curve. Due
to the doubling of transition paths, one may expect a strong
$ab$-plane spectral weight. However, by calculating the matrix
elements, the actual $ab$-plane spectrum is not so much weighted due
to the longer bonds in the $ab$-plane.  The shoulder in the
experimental E//$a$ curve at about 1.5 eV, {\it i.e.}, the energy of
peak $\alpha$ in the E//$c$ curve may originate from the spin
disordering in the $ab$-plane.  First we note that a very tiny peak is
seen at this energy in the C-AF ground state configuration.  This
structure mainly comes from the transition from the occupied
V-1-$d_{zx}^\uparrow$ component to a small V-2-$d_{zx}^\uparrow$
component just above the Fermi level.  However, this structure is too
weak to explain the shoulder at 1.5 eV.  If the spin state between V-1
and V-2 may become parallel like in A-AF and FM, the transition
corresponding to the red dashed arrow may become allowed and produces
a peak as shown in Fig.3.

From the above assignment, some important parameters can be estimated
by using the Hartree-Fock model. The peak position of $\alpha$ should
corresponds to energy $U-J$, and peak $\beta$ corresponds to $U+J$,
where $U$ and $J$ are on site Coulomb and exchange parameters. The
estimated values for $U$ and $J$ are 2.2eV and 0.6eV from our
calculations, and about 2.1eV and 0.32eV from experiments for
LaVO$_3$. The smaller $J$ parameter estimated from experiments may
suggests the spin canting or fluctuation at elevated temperature,
especially for YVO$_3$ as discussed below.

\subsubsection{YVO$_3$ at low temperature (T$<$77K):}

Now let us discuss the low temperature phase of YVO$_3$. In this case,
the C-OO with G-SO state is the most stable state. As the results, the
strong peak $\beta$ for E//$c$ can be explained as the transition from
V-1-$d_{yz}^\uparrow$ to V-3-$d_{yz}^\uparrow$ as noted by the solid
red arrow in Fig.2B with energy $U+J$. This transition is not allowed
for the FM coupling along the $c$-axis as in the case of C-AF or FM
state, and should show strong temperature dependence as observed in
experiment. As for the small peak $\alpha$, the situation for C-OO is
different from the case of LaVO$_3$ with G-OO.  In the present case of
C-OO, the significant mixture of different orbitals under strongly
distorted VO$_6$ environment is crucially important to produce this
small peak $\alpha$ located at energy $U-J$ (corresponding to the
transition as noted by the green arrow in Fig.2B).  Let us explain the
situation in more details.  If there is no orbital mixing (the
inter-site hybridization for the same orbitals still exists), $d_{yz}$
is fully occupied and $d_{zx}$ is empty in the majority spin state
both at V-1 and V-3. The unoccupied $d_{zx}^\uparrow$ at V-1
hybridizes with unoccupied $d_{zx}^\uparrow$ (not $d_{yz}^\uparrow$)
at V-3 to produce a small $d_{zx}^\uparrow$ component at V-3 site at
the energy of majority spin empty state.  In this case, the weight of
the transition from V-1-$d_{zx}^\uparrow$ to V-3-$d_{zx}^\uparrow$ is
zero because of the zero occupation of V-1-$d_{zx}^\uparrow$ state,
and the small peak $\alpha$ should not exist.  However, as already
mentioned the orbital mixing between $d_{yz}$ and $d_{zx}$ is
significant due to the lattice distortion and produces the partial
occupation of V-1-$d_{zx}^\uparrow$ state (green line in Fig.2B). This
will contribute to the small peak $\alpha$.  We should emphasize here
that by keeping the C-OO any magnetic coupling along the $c$-axis will
produce a peak around $\alpha$ only through the orbital mixing and
therefore the peak $\alpha$ will never be strong, in contrast to the
case of LaVO$_3$. Nevertheless, those states with different orbital
ordering, like the intermediate temperature phase with G-OO and C-SO,
can contribute to the peak structure around $\alpha$ as shown in
Fig.2C.

\subsubsection{YVO$_3$ at intermediate temperature (77K$<$T$<$116K):}

The observed spectra for the intermediate phase of YVO$_3$, which has
G-OO and C-SO, is hard to be explained by our calculations. Our
calculations suggest a picture similar to that of the low temperature
phase of LaVO$_3$, because the two systems have the same spin and
orbital orderings. However, the observed spectra is quite similar to
the low temperature phase of YVO$_3$, which has different spin and
orbital orderings. In particular, the observed peak $\beta$ is
stronger than peak $\alpha$ and temperature dependent, while the
calculations suggest the opposite. One possible scenario to explain
the observed picture is to suggest the persistence of low temperature
phase (C-OO and G-SO) at elevated temperature, or at least certain
mixture of other magnetic states, like G-OO with G-SO or G-OO with
A-SO states. This argument can be supported by the following
facts. First, the small $J$ parameter (about 0.25eV) estimated from
experiment suggests the possibility of strong canting or magnetic
fluctuation at elevated temperature (100K). Second, the recent neutron
experiment~\cite{Keimer} on this compound actually suggests the
existence of G-type spin canting and significant reduction of magnetic
moment at 100K. Nevertheless, according to our calculations, to be
able to reproduce the observed shape, substantial (more than 50
percent) mixture of low temperature phase is required.

In conclusion, we developed the first-principles
plane-wave-pseudopotential method based on LDA+$U$ to calculate the
inter-band optical conductivity. We calculated the anisotropic optical
conductivities of LaVO$_3$ and YVO$_3$ for different orbital and spin
ordered phases, and studied the effects of spin and orbital ordering on
the anisotropy. The experimentally observed spectra for both LaVO$_3$
and YVO$_3$ at low temperature can be qualitatively explained by our
calculations. On the other hand, our calculation for the 
intermediate temperature
phase for YVO$_3$ suggests that substantial mixture of 
low temperature phase may
persist at elevated temperature.

\begin{acknowledgments}
The authors appreciate Prof. Y. Tokura and Dr. S. Miyasaka for
fruitful discussions and providing their experimental data. The
authors also acknowledge the valuable discussion with
Dr. I. V. Solovyev and Dr. Y. Motome.
\end{acknowledgments}


\begin{thebibliography}{}
\bibitem{Orbital} Y. Tokura and N. Nagaosa, SCIENCE {\bf 288}, 462
(2000).
\bibitem{LSMO} Z. Fang, I. V. Solovyev and K. Terakura,
Phys. Rev. Lett. {\bf 84}, 3169 (2000).
\bibitem{JPC} Z. Fang and K. Terakura, J. Phys.: Condens. Matter {\bf
14}, 3001 (2002).
\bibitem{X-ray} Y. Murakami, J. P. Hill, D. Gibbs, and {\it et al.},
Phys. Rev. Lett {\bf 81}, 582 (1998); Y. Murakami, H. Kawada,
H. Kawata, and {\it et al.}, Phys. Rev. Lett. {\bf 80}, 1932 (1998).
\bibitem{PCMO} Y. Okimoto, Y. Tomioka, Y. Onose, and {\it et al.},
Phys. Rev. B {\bf 59}, 7401 (1999).
\bibitem{LMO} K. Tobe, T. Kimura, Y. Okimoto, and K. Tokura,
Phys. Rev. B {\bf 64}, 184421 (2001).
\bibitem{Millis} K. H. Ahn and A. J. Millis, Phys. Rev. B {\bf 61},
13545 (2000).
\bibitem{Igor} K. Terakura, I. V. Solovyev, and H. Sawada, in {\it
Colossal Magnetoresistive Oxides}, edited by Y. Tokura (Gordon \&
Breach Science Publishers, London, 2000).
\bibitem{LVO} P. Bordet, C. Chaillout, M. Marezio, and {\it et al.},
J. Soli. Stat. Chem. {\bf 106}, 253 (1993).
\bibitem{Miyasaka2} S. Miyasaka, T. Okuda, and Y. Tokura,
Phys. Rev. Lett. {\bf 85}, 5388 (2000).
\bibitem{YVO1} H. Kawano, H. Yoshizawa, and Y. Ueda,
J. Phys. Soc. Jpn. {\bf 63}, 2857 (1994).
\bibitem{YVO2} M. Noguchi, A. Nakazawa, S. Oka, and {\it et al.},
Phys. Rev. B {\bf 62}, R9271 (2000).
\bibitem{YVO3} G. R. Blake, T. T. M. Palstra, Y. Ren, and {\it et
al.}, Phys. Rev. Lett. {\bf 87}, 245501 (2001); {\it ibid},
Phys. Rev. B {\bf 65}, 174112 (2002).
\bibitem{YVO4} Y. Ren, T. T. M. Palstra, D. I. Khomski, and {\it et
al.}, NATURE {\bf 396}, 441 (1998).
\bibitem{sawada1} H. Sawada, and K. Terakura, Phys. Rev. B {\bf 53},
12742 (1996).
\bibitem{LDAU} V. I. Anisimov, J. Zaanen, and O. K. Anderson,
Phys. Rev. B {\bf 44}, 943 (1991); I. V. Solovyev, P. H. Dederichs,
and V. I. Anisimov, Phys. Rev. B {\bf 50}, 16861 (1994).
\bibitem{sawada2} H. Sawada, and K. Terakura, Phys. Rev. B {\bf 58},
6831 (1998).
\bibitem{Miyasaka} S. Miyasaka, Y. Okimoto, and Y. Tokura,
J. Phys. Soc. Jpn., (to appear); cond-mat/0207664.
\bibitem{PP} D. Vanderbilt, Phys. Rev. B {\bf 41}, 7892 (1990).
\bibitem{Kubo} C. S. Wang and J. Callaway, Phys. Rev. B {\bf 9}, 4897
(1974).
\bibitem{Fujiwara} T. Fujiwara, and T. Hoshi, J. Phys. Soc. Jpn. {\bf
66}, 1723 (1997).
\bibitem{Matrix} H. Kageshima, and K. Shiraishi, Phys. Rev. B {\bf
56}, 14985 (1997).
\bibitem{gap} T. Arima, Y. Tokura, and J. B. Torrance, Phys. Rev. B
{\bf 48}, 17006 (1993).
\bibitem{Mizokawa} T. Mizokawa, and A. Fujimori, Phys. Rev. B {\bf
54}, 5368 (1996); T. Mizokawa, D. I. Khomskii, and G. A. Sawatzky,
Phys. Rev. B {\bf 60}, 7309 (1999).
\bibitem{Keimer} C. Ulrich, G. Khaliulin, J. Sirker, and {\it et al.},
(to be published).

\end{thebibliography}
\end{document}